\documentclass[12pt]{article}
\usepackage[english]{babel}							
\usepackage[utf8]{inputenc}
\usepackage[T1]{fontenc}
\usepackage{latexsym,amsmath,amsfonts,amssymb,amsthm,cancel,siunitx,calculator,calc,mathtools,empheq}
\usepackage{subfig,epsfig,tikz,float}
\usepackage{booktabs,multicol,multirow,tabularx,array}
\usepackage{natbib}
\usepackage{setspace}
\usepackage{hyperref}
\usepackage{url}
\setcitestyle{numbers}
\setcitestyle{square}
\title{Scoring Aave Accounts for Creditworthiness}

\author{
Will Wolf \\ will@credprotocol.com \and
Aaron Henry \\ aaron@credprotocol.com \and
Hamza Al Fadel \\ hamza@credprotocol.com \and
Xavier Quintuna \\ xavier@credprotocol.com \and
Julian Gay \\ julian@credprotocol.com
}

\begin{document}
\sloppy

\date{}

\maketitle

\vspace{-0.5cm}

\bigskip
\noindent
{\small{\bf ABSTRACT.}
Scoring the creditworthiness of accounts that interact with decentralized financial (DeFi) protocols remains an important yet unsolved problem. In this paper, we propose a credit scoring system for those accounts that have interacted with the Aave v2 liquidity protocol. The key component of this system is a tree-based binary classifier that predicts ``position delinquency.'' To the community, we provide our method, results, and the (abridged) dataset on which this system is built.
}


\baselineskip=\normalbaselineskip

\section{Introduction}\label{sec:1}

Credit scoring is the task of quantifying the creditworthiness of a given account. In traditional finance, this ``account'' is a legal entity, e.g. a person or business, and ``creditworthiness'' is proxied by this entity's propensity to repay a loan outstanding. Typically, the former is identified by a government identification number, e.g. a social security number (SSN) or employer identification number (EIN) in the United States, while the latter is presented as an integer ``credit score,'' e.g. that of TransUnion \cite*{transunion2022}, Experian \cite*{experian2022}, or Equifax \cite*{equifax2022}. Accounts with a high credit score enjoy numerous financial advantages—favorable terms on credit cards, loans, mortgages, property rental, insurance, etc.—and vice versa. In this vein, credit scores enable \emph{risk-adjusted} pricing—facilitating more stable, predictable, and \emph{capital-efficient} financial services writ large.

``Decentralized finance'' (DeFi) recapitulates many of these financial services, such as trading, insurance, and lending. However, as no salient system of credit scoring currently exists, a substantial proportion of DeFi services remain glaringly \emph{capital-inefficient}. For example, ``on MakerDAO, borrowers are required to collateralize their loan with, at a minimum, 150\% of the loan value [...] most individuals will collateralize their loans well over 200\%, with the average collateralization ratio across all platforms being 348\%'' \cite*{defirate2022} as of September 2021. As such, building a transparent, effective, and statistically-defensible DeFi credit score remains a critically important problem to solve.

\begin{figure}[!htb]
\centering
\includegraphics[width=\textwidth]{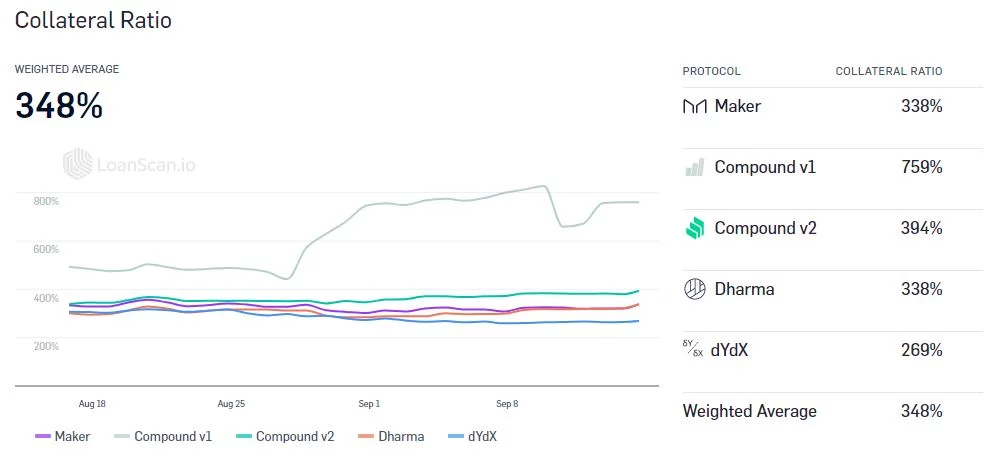}
\caption{Collateralization rates over time on various liquidity protocols}\label{fig:1}
\end{figure}

In this paper, we propose a system for scoring the creditworthiness of those Ethereum ``accounts''—unique 64-byte addresses on the Ethereum blockchain—that have interacted with Aave v2—a popular DeFi ``liquidity protocol'' used for borrowing and lending cryptocurrency. In addition, we detail and make public the (abridged) dataset on which this system is built. Presently, this system is running in production on beta.credprotocol.com.

\section{Related Work}\label{sec:2}

In traditional finance, the typical credit scoring model is a binary classifier trained to predict loan repayment delinquency. Concretely, given historical account behavior—features $\mathbf{X}$ detailing, for instance, ``payment history, amounts owed, length of credit history, new credit, and credit mix'' \cite*{myfico2022}—the model predicts the probability $\tilde{y} = p(y\vert\mathbf{X})\in [0, 1]$ that an account will repay a given loan in the 90 days following its repayment date \cite*{fedres2007}. Here, $y = 1$ denotes delinquency, termed a ``bad'' account, while $y = 0$ denotes repayment, termed a ``good'' account. This prediction $\tilde{y}$ is then mapped to an integer credit score, e.g. $s \in \{k \in \mathbb{N} : 300 \leq k \leq 850\}$ in the case of FICO \cite*{goodcreditscore2022}, then presented to the end-user. Broadly, a higher score implies a higher likelihood of the account being ``good'' and vice versa. Such credit scoring systems have been in use for several decades to great effect in countries around the world.

At the time of writing, numerous web3 firms are competing to build and/or provide a reliable DeFi credit score: RociFi \cite*{rocifi2022}, ARCx \cite*{arcx2022}, CreDA \cite*{creda2022}, TRAVA \cite*{trava2022}, Quadrata \cite*{quadrata2022}, Credefi \cite*{credefi2022}, Spectral \cite*{spectral2022}, TrueFi \cite*{truefi2022}, Telefy \cite*{telefy2022}, Livesight \cite*{livesight2022}, and Masa \cite*{masa2022}, among others. Per Spectral's documentation, they train a tree-based classifier to predict ``whether a borrower has gotten liquidated within a predefined time window post the date of borrowing and whether his health factor dropped below a certain threshold within the same time frame'' using features detailing the account's ``transaction history, liquidation history, amounts owed and repaid, credit mix, and length of credit history.'' These scores are then ``scaled to the final score range of 300 to 850.'' Similarly, TrueFi ambitions to build ``a creditworthiness score from 0 to 255'' for crypto-native \emph{institutions} given features detailing ``company background, repayment history, operating and trading history, assets under management, credit metrics.'' At the time of writing, to the best of our knowledge, the remaining competitors do not document their approaches publicly.

\section{Scoring Aave Accounts}\label{sec:3}

The traditional credit scoring model formulation—a binary classifier trained to predict loan repayment delinquency—implies at least the following two assumptions:

\begin{enumerate}
    \item Atomic loans, i.e. the ability to distinguish two loans as distinct and mutually exclusive, exist.
    \item Discrete notions of atomic loan ``repayment,'' e.g. ``loan repaid in full in less than 90 days'' or ``loan repaid in part in more than 180 days,'' exist.
\end{enumerate}

In stark contrast, the Aave protocol has no notion of an atomic loan; instead, each account holds a diverse \emph{position}: a distinct counter of assets borrowed and designated as collateral at a given point in time. Unfortunately, this violates the first assumption above—which immediately violates the second.

\subsection{Health Factor}

To similar effect, account ``standing'' on Aave is characterized by a single, aggregate statistic, termed ``health factor'' ($\text{HF}$) \cite*{riskparams2022}, given by the following formula:

$$
\text{HF} = \frac{\sum\limits\text{Collateral}_i \text{ in ETH} *\text{Liquidation Threshold}_i}{\text{Total Borrows in ETH}}
$$

Here, $i$ represents a specific collateral asset, and its ``liquidation threshold'' the maximum fractional value of allowable corresponding debt. For example, in a position with a single collateral asset $c$ worth 10 ETH, and a liquidation threshold of $\text{LT}_c = 0.8$, the allowable value (in ETH) of debt outstanding at any given time is $\leq$ 8 ETH.

Crucially, should a position's HF fall below 1—due to a decrease in the value of its collateral or an increase in the value of its debt—it is eligible for $\emph{liquidation}$. In this event, a \emph{liquidator} may repay ``up to 50\% of the account's debt in exchange for an equivalent amount of its collateral \emph{plus a bonus}.'' \cite*{liquidations2022} Though liquidation is by design economically advantageous for a liquidator, it is performed at the latter's discretion. In this sense, an Aave position is not ``repaid''; instead, it is eligible for liquidation ($\text{HF} < 1)$, implying delinquency, or not ($\text{HF} \geq 1)$, implying responsible financial behavior. Trivially, we denote the former case as ``bad'' and the latter as ``good.''

\subsection{Problem formulation}

In the spirit of traditional financial credit scoring models, we seek to capture the probability with which an account—given its present and historical on-chain financial behavior—would fail to ``repay'' a subsequent loan. In this vein, for Aave accounts, we propose to model the probability that a novel position will become eligible for liquidation in its first 90 days. Here, define the discrete set of assets (tokens) transacted on Aave as $\mathcal{A} = \{\text{ETH, LINK, DAI, USDC, ...}\}$ with $|\mathcal{A}| \approx 35$ \cite*{risksperasset2022}; the counter of borrowed assets as $\mathcal{B} = \{(a, \#)_b : \forall{a \in \mathcal{A}}; \# \in \mathbb{R}_{\geq 0}\}$; and the counter of collateral assets as $\mathcal{C} = \{(a, \#)_c : \forall{a \in \mathcal{A}}; \# \in \mathbb{R}_{\geq 0}\}$. Next, define the account's position as $\mathcal{P}_i = \mathcal{B} \cup \mathcal{C}$, with $i$ representing the position's \emph{temporal index} in contiguous time. For instance, imagine the following sequence of positions for account $a$ and times $t$:

\begin{align*}
    \mathcal{P}_0 = \{(\text{DAI}, 1000)_c, (\text{ETH}, 0.1)_b\} \qquad (t = 0)\\
    \mathcal{P}_0 = \{(\text{DAI}, 1000)_c, (\text{ETH}, 0.1)_b\} \qquad (t = 1)\\
    \mathcal{P}_1 = \{(\text{DAI}, 1500)_c, (\text{ETH}, 0.1)_b\} \qquad (t = 2)\\
    \mathcal{P}_1 = \{(\text{DAI}, 1500)_c, (\text{ETH}, 0.1)_b\} \qquad (t = 3)\\
    \mathcal{P}_1 = \{(\text{DAI}, 1500)_c, (\text{ETH}, 0.1)_b\} \qquad (t = 4)\\
    \mathcal{P}_2 = \{(\text{DAI}, 1000)_c, (\text{ETH}, 0.1)_b\} \qquad (t = 5)
\end{align*}

Although positions $\mathcal{P}_0$ and $\mathcal{P}_2$ are \emph{equivalent}, we denote them differently, as they are ``opened'' at different points in time. Finally, define the target $y$, position delinquency, as $y = \text{max}\big(\{\text{HF}_{t|\mathcal{P}_i} < 1: t|\mathcal{P}_i \leq 90\}\big) \in \{0, 1\}$, with $t$ measured in days. Finally, we seek to model $p(y\vert\mathbf{X}_{t = 0|\mathcal{P}_i})$, where $\mathbf{X}_{t = 0|\mathcal{P}_i}$ are features detailing both the account's historical on-chain behavior up to and including the block in which it opens $\mathcal{P}_i$ as well as information about $\mathcal{P}_i$ itself. Contrary to Spectral, we do not consider whether the position is \emph{actually} liquidated; given $\text{HF} < 1$, account features $\mathbf{X}_{t = 0|\mathcal{P}_i}$ bear no additional influence on, i.e. are conditionally independent of, the liquidation event.

\section{Method}\label{sec:4}

\subsection{Aave v2 Health Factor Dataset}

The Aave v2 Health Factor Dataset contains an account's position and health factor information at 15-minute intervals. It is generated using historical token prices, aToken reserve specifications, and liquidity index references. We build our model features and target from this dataset.

To the community, we provide an abridged version of this dataset at \url{https://github.com/credprotocol/Aave-v2-Health-Factor-Dataset}. This version contains identical data yet on 1-week intervals. Its repository includes download instructions, as well as further information regarding its provenance, contents, and quality.

\subsection{Model}

On the \text{full} version of our dataset, we train a binary classifier $\mathcal{C}$ to predict $\hat{y} = p(y\vert\mathbf{X}_{t = 0|\mathcal{P}_i})$. Presently, $\mathbf{X}_{t = 0|\mathcal{P}_i}$ details the corresponding account's age, aggregations of the time series of its historical health factors, its interactions with the Aave protocol, the types of assets it borrows and keeps as collateral, and more. Before training, we remove all ``short-term'' positions—defined as those kept for $< 10$ days. Empirically, we find that a large majority of short-term positions resemble flash loans effectuated by smart contracts (used as part of high-volume arbitrage strategies) as opposed to the account itself. Conversely, long-term positions more closely resemble ``genuine'' borrowing behavior. In this study, we model the latter only. 

After applying this filter, we train on all account positions \emph{excluding} those \emph{current}. We then predict on the latter subset to obtain account-level delinquency scores.

\subsection{Evaluation}

We evaluate our model by measuring the area under the receiver operating curve (AUC). This statistic quantifies, classification-threshold irrespective, the model's ability to score positive examples ($y = 1$) more highly than negative examples ($y = 0$). 

In addition to our proposed model, we evaluate the following baseline models; for some account $a$, predict $\hat{y}$ as:

\begin{enumerate}
    \item A random draw from $\text{Uniform}(0, 1)$.
    \item $a$'s historical delinquency frequency (over all positions preceding the test example).
    \item $a$'s historical count of blocks in which $\text{HF} < 1$ (over all blocks preceding that of the test example).
\end{enumerate}

Finally, for our proposed model, we consider both logistic regression and tree-based classifiers. Respectively, in order, these 5 models are termed: ``random,'' ``E(y)\_lagged,'' ``count\_HF\_lt\_1\_lagged,'' ``log-reg,'' and ``tree-based.''

Our training dataset consists of roughly 34,000 rows. To compute AUC scores, we ``hold out'' the most recent (by block timestamp) 2,500 rows; this way, we evaluate our model's ability to score future positions. Then, we split these rows into 5 chunks of 500 rows each. Next, we train on the first 31,500 rows, predict on the first holdout chunk; train on the first 32,000 rows, predict on the second holdout chunk; etc. This gives ``out of fold'' predictions for all $500 * 5 = 2,500$ test rows for each model variant. The respective ROC curves and AUC scores are shown in Figure 2.

\begin{figure}[!htb]
\centering
\includegraphics[width=\textwidth]{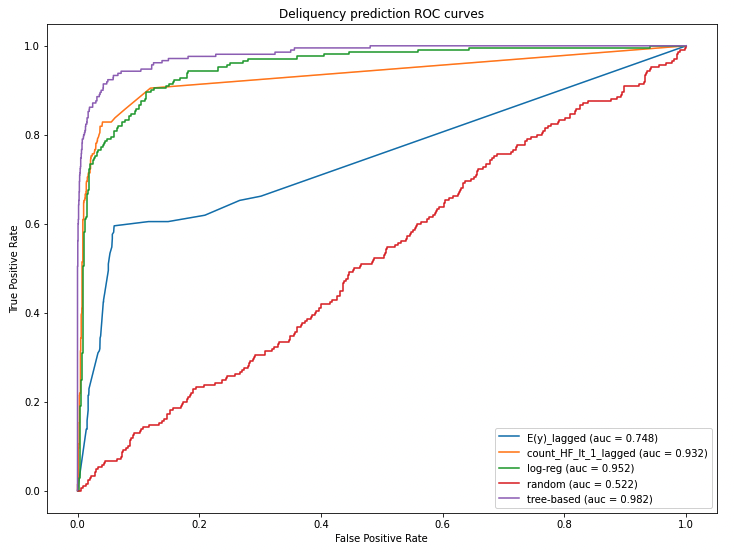}
\caption{ROC curves for proposed model and several baselines}\label{fig:2}
\end{figure}

From this plot, we determine that our ``tree-based'' classifier (shown in purple) is a strong predictor of position delinquency (and by proxy, creditworthiness). 

In addition, we highlight that our single-feature ``count\_HF\_lt\_1\_lagged'' baseline (shown in orange) is highly competitive itself (AUC: 0.932). Logically, this makes sense for at least the following three reasons:

\begin{enumerate}
    \item Accounts can employ smart contracts to programmatically ``rebalance'' their position when their health factor falls below a given threshold. For instance, by adding more collateral, or paying off debt. In this way, an account can significantly reduce the probability that its position ever falls delinquent.
    \item Holding stablecoins (e.g. DAI, USDC, etc.) as debt or collateral (or both) significantly reduces the variance of one's health factor.
    \item Broadly, past behavior is a reliable predictor of future behavior.
\end{enumerate}

As such, given at least a faithful historical time series of account health factors, one can build a compelling model of ``long-term'' position delinquency.

\subsection{Quantile Transform}

Our final credit score is an integer $s$ in the closed interval $[300, 1000].$ In order to map predictions $\hat{y}$ to this space, we first define a transformation operator $g$ and target distribution $p^*$, where $g: \hat{y} \sim p(y\vert\mathbf{X}_{t = 0|\mathcal{P}_i}) \rightarrow s \sim p^*$. In our current system, $p^*$ is made to (approximately) resemble the distribution of empirical FICO scores \cite*{ucr2005} given in Figure 3.

\begin{figure}[!htb]
\centering
\includegraphics[width=\textwidth]{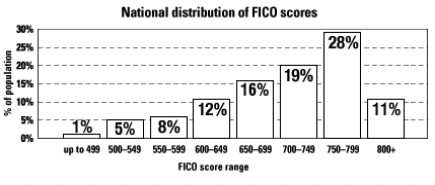}
\caption{Empirical distribution of FICO scores}\label{fig:3}
\end{figure}

To distribute our scores as such, we first scale the ``FICO score range'' bins to fit our stated interval, then simulate points from the now ``stretched,'' pseudo-empirical probability mass function given above. Next, we fit the location, scale, and shape parameters of a skew-normal distribution to these samples. Lastly, we use $g$ to ``quantile-transform'' our predictions into the target distribution $p^*$, then round down to the nearest integer. This gives our final credit (Cred) score.

For Aave accounts, our scores are distributed as shown in Figure 4.

\begin{figure}[!htb]
\centering
\includegraphics[width=\textwidth]{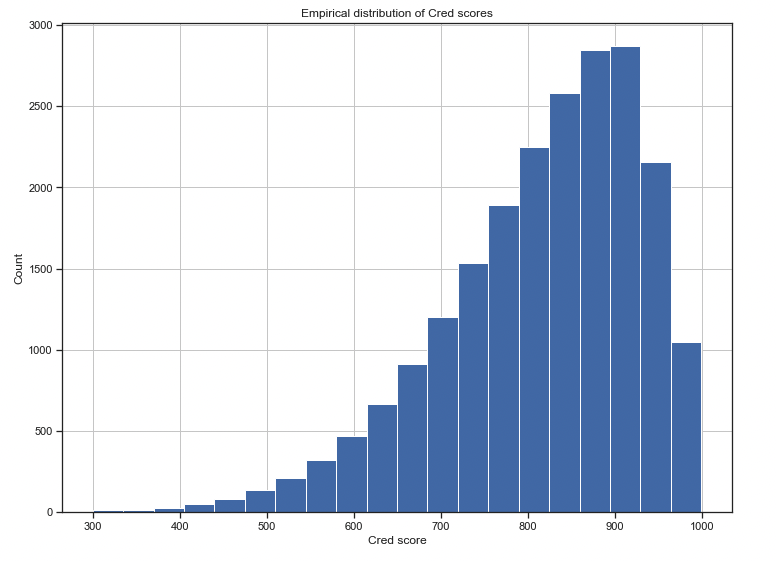}
\caption{Empirical distribution of Cred scores}\label{fig:4}
\end{figure}

Additionally, Figure 5 shows the empirical mapping between delinquency probability and integer Cred score.

\begin{figure}[!htb]
\centering
\includegraphics[width=\textwidth]{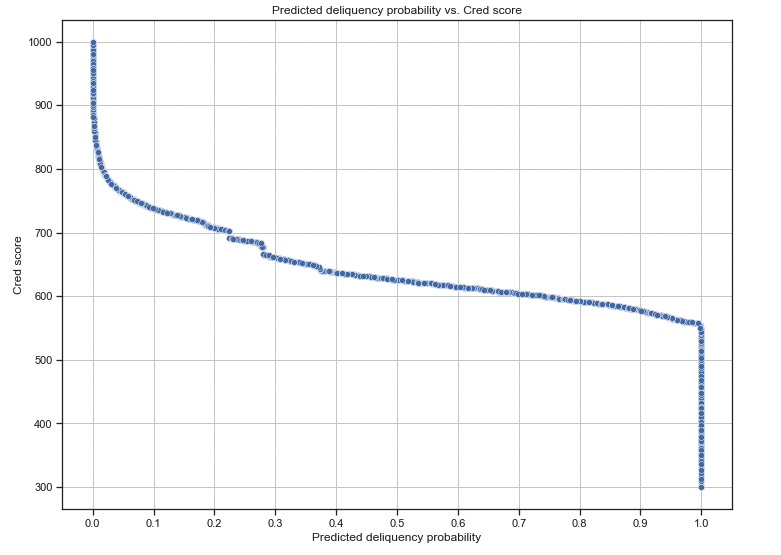}
\caption{$p(y\vert\mathbf{X}_{t = 0|\mathcal{P}_i})$ vs. Cred score}\label{fig:5}
\end{figure}

\section{Discussion}\label{sec:5}

Performing dynamic and accurate DeFi credit scoring represents an enormous opportunity for our space. As Naeem Siddiqi, Director of Credit Scoring and Decisioning with SAS Institute, notes in his seminal text \cite*{siddiqi2017}, credit scoring at once ``provides lenders with an opportunity for consistent and objective decision making, based on empirically derived information''; ``makes it easier for good customers to access credit as they now have strong, reliable evidence of their satisfactory payment behavior''; and so on. In applications built on blockchains—financial and non-financial alike—credit scoring models might be used to (first and foremost) enable undercollateralized DeFi lending; qualify loan applicants; enable risk-based portfolio construction; assess the fidelity of DAOs, multi-sig wallets, or DeFi protocols themselves per the creditworthiness of their members; intrinsically motivate constructive financial behavior; and so on. Furthermore, building this system from on-chain data allows us to update these scores as quickly as new blocks are added to the blockchain.

In this paper, we've presented a credit scoring system built from on-chain account interactions with the Aave v2 liquidity protocol. These scores currently power Cred Protocol's various data products. In the future, we look forward to applying similar methodologies to accounts on other liquidity protocols like Compound, MakerDAO, and more. To the community, we welcome earnest feedback, collaboration, and further research in this space.

\section{Future Work}\label{sec:6}

The system presented in this paper predicts 90-day delinquency of ``long-term'' positions. Nominally, the impact of market volatility on an account's health factor—as a function of relative changes in collateral and debt asset values—should impact this outcome. While current market prices are considered in our model, historical nor simulated (future) prices are not.

One alternative model might consider the account as a \emph{stochastic inventory controller} which, in order to maintain $\text{HF} \geq 1$, continuously adjusts its position to account for past and (predicted) future market movement. In this vein, and assuming a current position $\pi = \mathcal{P}_i$, we might model $p\big(\Delta\pi_{t \rightarrow t + 1} | (\Delta\text{HF}_{t \rightarrow t + \epsilon} | \pi_t; \mathcal{M}_{t + \epsilon})\big)$, which answers the following question: given the impact of instantaneous market conditions $\mathcal{M}_{t + \epsilon}$ on health factor, how does the account \emph{modulate} its position in turn? Next, conditional on historical asset prices, we might simulate future asset price trajectories via a (correlated, multi-asset) Brownian motion process, then employ this controller at each time step to predict $\Delta\pi_{t \rightarrow t + 1}$, and therefore $\text{HF}_{t + 1}$, outright. Flexibly aggregating this predicted sequence over arbitrary future time windows—for example, by computing the proportion of its elements that fall below 1—is then trivial.

Separately, as updates to our present system, we may also wish to predict ``short-term'' position delinquency, and potentially over a future window of \emph{user-specified} length. To the former, we might add additional features regarding the position-effectuating \emph{transaction}; for instance, the type of signer, the specific steps involved, or perhaps even a hash of the bytecode itself. To the latter, we might train our model to predict $p(y\vert\mathbf{X}_{t = 0|\mathcal{P}_i}; L)$, where $y$ denotes delinquency in the subsequent $L$ days. During training, one could create an arbitrary number of examples of varying $L$; at test time, $L$ is provided by the user.

We leave both directions to future work.

\section{Acknowledgements}\label{sec:7}

We thank Michael Griffiths, Nathan Matare, and Brandon Rochon for their thoughtful feedback and review.

\newpage

\bibliographystyle{apalike}
\bibliography{ref.bib}

\end{document}